\documentclass{aa}  
\usepackage{graphicx}
\usepackage{amssymb}
\begin{document}
\title{Wide and deep near-UV (360nm) galaxy counts and the extragalactic
   background light with the Large Binocular Camera}

\author{
          A. Grazian \inst{1}
          \and
          N. Menci \inst{1}
          \and
          E. Giallongo \inst{1}
          \and
          S. Gallozzi \inst{1}
  	  \and
	  F. Fontanot \inst{2,3}
          \and
          A. Fontana \inst{1}
          \and
          V. Testa \inst{1}
	  \and
          R. Ragazzoni \inst{4}
          \and
          A. Baruffolo \inst{4}
          \and
          G. Beccari \inst{5}
          \and
          E. Diolaiti \inst{6}
          \and
          A. Di Paola \inst{1}
          \and
          J. Farinato \inst{4}
          \and
          F. Gasparo \inst{2}
          \and
          G. Gentile \inst{4}
          \and
          R. Green \inst{7}
          \and
          J. Hill \inst{7}
          \and
          O. Kuhn \inst{7}
          \and
          F. Pasian \inst{2}
          \and
          F. Pedichini \inst{1}
          \and
          M. Radovich \inst{8}
          \and
          R. Smareglia \inst{2}
          \and
          R. Speziali \inst{1}
          \and
          D. Thompson \inst{7}
          \and
          R. M. Wagner \inst{7}
}

   \offprints{A. Grazian, \email{grazian@oa-roma.inaf.it}}

\institute{INAF - Osservatorio Astronomico di Roma, Via Frascati 33,
I--00040, Monteporzio, Italy
\and INAF - Osservatorio Astronomico di Trieste, Via G. B. Tiepolo 11,
I--34131 Trieste, Italy
\and MPIA Max-Planck-Institute f\"ur Astronomie, K\"onigstuhl 17, D--69117
Heidelberg, Germany
\and INAF - Osservatorio Astronomico di Padova, vicolo dell'Osservatorio 5,
I--35122 Padova, Italy
\and ESA, Space Science Department, 2200 AG Noordwijk, Netherlands
\and INAF - Osservatorio Astronomico di Bologna, Via Ranzani 1,
I--40127 Bologna, Italy
\and Large Binocular Telescope Observatory, University of Arizona, 933 N.
Cherry Ave., Tucson, AZ 85721-0065
\and INAF - Osservatorio Astronomico di Capodimonte, via Moiariello 16,
I--80131, Napoli, Italy
}

   \date{Received 20 March 2009; accepted 15 June 2009}

   \authorrunning{Grazian et al.}
   \titlerunning{Wide and deep near-UV galaxy counts with LBC}

  \abstract
   {Deep multicolour surveys are the main tool to explore the formation and
evolution of the very faint galaxies which are beyond the
spectroscopic limit with the present technology. The photometric
properties of these faint galaxies are usually compared with current
renditions of semianalytical models to provide constraints on the
detailed treatment of the fundamental physical processes involved in
galaxy formation and evolution, namely the mass assembly and the star
formation.}
   {Galaxy counts over large sky areas in the
360nm near-UV band are particularly important because they are difficult
to obtain given the low efficiency of near-UV instrumentation, even at 8m
class telescopes. Observing in the near-UV bands can provide a first guess
on the distribution of star formation activity in distant (up to $z\sim 3$)
galaxies. A relatively large instrumental field of view helps in minimizing the
biases due to the cosmic variance.}
   {We have obtained deep images in the 360nm U band provided by the blue
channel of the Large Binocular Camera at the prime focus of the Large
Binocular Telescope. We have derived over an area of $\simeq 0.4$ sq.
deg. the galaxy number counts down to $U=27$ in the Vega system
(corresponding to U=27.86 in the AB system) at a completeness level of
30\% reaching the faintest current limit for this wavelength and sky
area.}
   {The shape of the galaxy number counts in the U band can be described
by a double power-law, the bright side being consistent with the shape
of shallower surveys of comparable or greater areas. The slope bends
over significantly at $U>23.5$ ensuring the convergence of the contribution
by star
forming galaxies to the Extragalactic Background Light in the near-UV band
to a value which is more than 70\% of the most recent upper limits
derived for this band. We have jointly compared our near-UV and K
band counts collected from the literature with few selected
hierarchical CDM models emphasizing specific critical issues in the
physical description of the galaxy formation and evolution.}
   {}


   \maketitle
%

\section{Introduction}

Wide and deep multicolour surveys are useful tools to investigate in
detail the processes of galaxy formation and evolution, especially
beyond the spectroscopic capabilities
of current instrumentation. One of the main aims of the deep
multicolour surveys is to provide a clear picture of the processes
involved in the mass assembly and star formation of galaxies across
the cosmic time. Galaxy observables, e.g. luminosity and mass
functions, two point correlation function, are typically compared with
current renditions of semi-analytical models in hierarchical CDM
scenarios, in order to derive the key ingredients related to the
physics of galaxy formation
(\cite{croton06,menci06,bower06,nagamine,ff07,dlb07,somerville08,keres,dekel}).

The observed properties of the galaxy population, however, are mainly
affected by the limited statistics. Deep pencil beam surveys, such as
GOODS (\cite{goods}) or HUDF (\cite{udf}), have been carried out on
relatively small sky areas and are subject to the cosmic variance
effect. Larger surveys like COSMOS which extends over a 2 deg$^2$ area
are indeed shallower and limit the knowledge of the faint galaxy
population at intermediate and high redshifts.

To probe the statistical properties of the faint galaxy population
reducing the biases due to the cosmic variance, a major effort should
be performed over large areas with efficient multicolour imagers at 8m
class telescopes. This is especially true in the near-UV band (hereafter UV)
where instrumentation is in general less efficient but where it is possible
to extract information on the star formation activity and dust
absorption present in distant galaxies.

In this context we are exploiting the unique power of the Large
Binocular Camera (LBC) installed at the prime focus of the Large
Binocular Telescope (LBT,
\cite{pedik,speziali,ragazzoni06,hill,giallongo}) to reach faint
magnitude limits in the U band ($\lambda\sim 3600$\AA) over areas of
several hundreds of sq. arcmin.

Long LBC observations in the UV band are of comparable depth to those
of the HDFs (although, obviously, the image quality will be
poorer). Even a single pointing with LBC produces a field two
orders of magnitude larger than that of the combined HDF-N and
HDF-S. This is very important because the transverse extent of the
HDFs corresponds to about 1 Mpc at $z\sim 0.5-2$ where Dark Matter
clustering is still important.

The goal of this paper is to provide UV (360nm) galaxy number counts
down to the faintest magnitude limits available from ground based
observations. The comparison of normalization and shape of the
observed counts with that predicted by theoretical hierarchical models
can help to enlighten critical issues in the description of galaxy
formation and evolution like dust extinction and the formation of
dwarf galaxies.

Throughout the paper we adopt the Vega magnitude system
($U_{Vega}=U_{AB}-0.86$) and we refer to differential number counts
simply as ``number counts'', unless otherwise stated.


\section{The Data}

The deep UV observations described here have been carried out with the
Large Binocular Camera (LBC, \cite{giallongo}). LBC is a double imager
installed at the prime foci stations of the 8.4m telescopes LBT (Large
Binocular Telescope, \cite{hill}). Each LBT telescope unit are
equipped with similar prime focus cameras. The blue channel (LBC-Blue)
is optimized for imaging in the UV-B bands and the red channel
(LBC-Red) for imaging in the VRIZY bands. The unvignetted FoV of each
camera is 27 arcminutes in diameter, and the detector area is
equivalent to a $23\times 23$ arcmin$^2$ field, covered by four 4K by
2K chips of pixel scale 0.225 arcsec. Because the mirrors of both
channels are mounted on the same pointing system, a given target can
be observed simultaneously over a wide wavelength range, improving the
operation efficiency. Extensive description of the twin LBC instrument
can be found in \cite{giallongo,ragazzoni06,pedik,speziali08}.

We have used a deep
U-BESSEL image of 3 hours, acquired in normal seeing condition
(FWHM=1arcsec) during the commissioning of the LBC-Blue camera, to derived
faint UV galaxy-counts in a 478.2 $arcmin^2$ sky area till
U(Vega)$=26.5$, in the Q0933+28 field (\cite{steidel03}).
The data have been reduced using the LBC pipeline described in
detail in \cite{giallongo}, applying the standard debias, flat-fielding
and stacking procedures to derive the coadded image. The flux calibration
of the U-BESSEL image has been
derived through observations of photometric standards from the fields
SA98 and SA113 (\cite{landolt92}) and the photometric fields
of \cite{galadi00}, as described in detail in \cite{giallongo}.
The precision of the zero point calibration in the U-BESSEL filter
is typically of the order of 0.03 mag at 68\% confidence level.
A correction to the photometric zero point of 0.11 mag due to the
Galactic extinction has been applied to the final coadd U-BESSEL image.

The Q0933+28 field was also imaged in the SDT-Uspec\footnote{this is an
interference filter with a wavelength range similar to U-BESSEL filter, but
the peak transmission is $\sim 30\%$ more efficient than that, as
described in Fig.2 of \cite{giallongo}.} filter of LBC for
an additional hour in the first quarter of 2007, under normal seeing
conditions (1.1 arcsec). These images are reduced and coadded in the same
way of the U-BESSEL ones, except for the photometric calibration procedure,
that is carried out using a spectrophotometric standard star of \cite{oke}.
The precision of the zero point calibration in the SDT-Uspec filter
is typically of the order of 0.05 mag at 68\% c.l.
A Galactic extinction correction of 0.13 has been used for this image.

To push deeper the galaxy number counts in
the UV band, we sum up the image obtained in the U-BESSEL filter (with
exposure time of 3 hours) with this new one, after rescaling the two
images to the same zero point. We have verified that the effective
wavelengths of these two filters are the same, the only difference
being the higher transmission efficiency (1.5 times, after integrating its
efficiency curve from $\lambda$=3000 to 4000 \AA) of the SDT-Uspec
filter compared to the U-BESSEL one. The colour term between these two
filters is 0.01, thus we neglect it when summing up the two
images. The resulting image goes $\sim$0.5 mag deeper than the original 3
hours with U-BESSEL filter, given the higher image quality of the new
SDT-Uspec image due to the general improvement of the telescope-instrument
system, in particular the reduction of scattered light from the telescope and
dome environment after the first run of the LBC-Blue commissioning.
We use this final coadded image to improve the
magnitude limit in the UV band and extend the number counts in this
band to faint fluxes.

To decrease the effects of cosmic variance in the number counts at
$U\sim 20$ we have used 3 additional LBC fields with shallower magnitude limits
$U\le 25$ but much larger area (892 $arcmin^2$) in the Subaru XMM Deep
Survey (SXDS, \cite{sxds}) region. These images have been reduced and
calibrated as described above for the Q0933+28 field.
The FWHM of these images (SXDS1, SXDS2, SXDS3) is higher (1.2, 1.25, 1.4
arcsec) than the one in the Q0933+28 field and the exposure
time per LBC pointing is 1.0, 1.5, and 1.5 hours, respectively,
in the U-BESSEL filter.
These three LBC images, combined with the deep point in the Q0933+28 field
are then used to derive wide and deep galaxy number counts in the U band,
from $U=19.5$ to 25.0 over a FoV of 1370 $arcmin^2$ (0.38 sq. deg.)
and to $U=27.0$ for a sub-sample of $\sim$ 480 $arcmin^2$.


\section{The number counts in the U band}

\subsection{Deep U band galaxy number counts}

We computed galaxy number counts using the SExtractor package
(\cite{sex}). For objects with area greater than that
corresponding to a circular aperture of radius equal to the FWHM, we
used the ``best'' photometry (Kron magnitude or corrected
isophotal magnitude if
the galaxy is severely blended with surrounding objects) provided by
SExtractor. For smaller sources we computed magnitudes in circular
apertures with diameter equal to 2 times the FWHM, and correct them
with an aperture correction we derived using relatively bright stars
in the field.  This allows us to avoid the well known underestimate of
the flux of faint galaxies provided by the isophotal method. To
isolate the few stars from the numerous faint galaxies in this field,
we relied on the class\_star classifier provided by SExtractor. It is
known that the morphological star/galaxy classifier of SExtractor is
not reliable for faint objects, but the contamination from stars at
faint U band fluxes and at high galactic latitude is very limited
(1\%), according to the model of \cite{bs1980}. At brighter magnitudes,
$U\sim 20$, the contamination from stars cannot be neglected, but at
high S/N ratio the morphological classificator of SExtractor is
robust. Moreover, the agreement between the LBC number counts at $U\le 22$
with those of SDSS and other large area surveys (e.g., VVDS) indicates
that the contamination from galactic stars are reduced also for bright U band
magnitudes.

We search for an optimal configuration of the detection parameters of
SExtractor in order to maximize the completeness at faint magnitude
limits and to reduce the number of spurious sources. The two
parameters regulating the depth, completeness and reliability of
photometric catalog are the threshold (relative to the RMS of the
image) adopted for detection of objects and the minimum area of
connected pixels above this threshold. We used the negative image
technique, described in \cite{dickinson2004,yw04,bouwens07}, as an
estimate of the reliability of the catalog. Using SExtractor, we
produce the ``-OBJECTS'' image, i.e. the image with the detected
objects subtracted, then we compute the negative of this image and run
SExtractor with the same detection parameters used for the positive
image. The sources detected on the negative image give an estimate of
the contamination from spurious sources, provided that the noise
statistic is symmetric (towards positive and negative pixel values) in
the image.

After several tests with different parameter set, the best effort
combination are thresh=0.7, area=8 pixels for Q0933+28
field and thresh=0.8, area=9 for SXDS fields (where the seeing is a bit
worse), which minimize the
number of spurious sources (measured in the negative image) while
maintaining the number counts and completeness of real sources
relatively high at faint magnitude limits ($U\sim 26.5$). In the
Q0933+28 field the contamination from spurious sources is $\sim 2.2\%$
at $U=26.5$, and remains less than $2.5\%$ till $U=27.0$
(corresponding to U(AB)=27.9).

The results on the galaxy number count analysis can be affected by
blending of galaxies when deep fields are investigated. In SExtractor
the two parameters affecting the deblending of sources are the {\sc
DEBLEND\_NTHRESH} and {\sc DEBLEND\_MINCONT} parameters. We produce
different catalogs using various combination of these two parameters
for each LBC image, and we find that the raw number counts are
not sensitive to these parameters, since the number of galaxies per
magnitude bin and per square degree varies of $\Delta Log N=0.02$ at
$U=24-25$ and $\Delta Log N=0.04$ at $U=27$, our faintest bin in the
galaxy number counts. We adopt {\sc DEBLEND\_NTHRESH=32} and {\sc
DEBLEND\_MINCONT=0.002} for all the fields described here, based on
visual inspection of the reliability of deblended sources on the
Q0933+28 field.

\begin{figure}
\includegraphics[width=9cm]{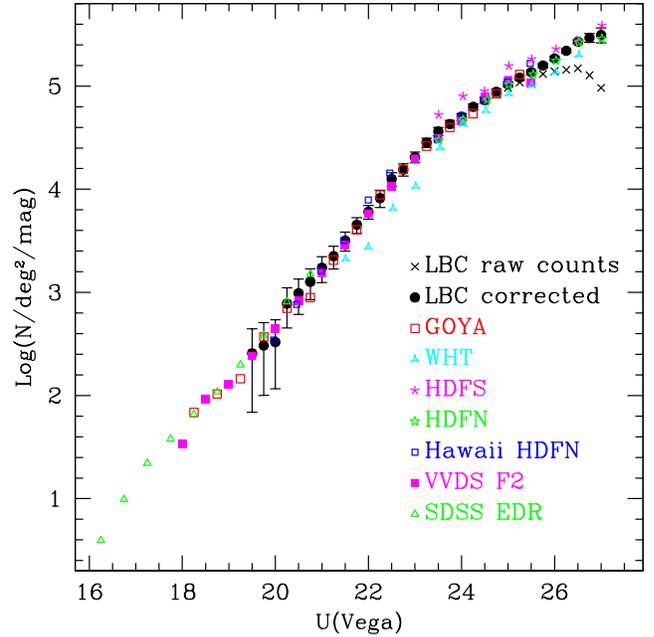}
\caption{
Number counts of galaxies in the U-BESSEL band for the Q0933+28 and SXDS
LBC fields. Black crosses represent the raw galaxy number counts, while
filled big circles show the number counts corrected for incompleteness.
Magnitudes are in the Vega system. We compare our counts with
shallow surveys of similar or larger area (SDSS EDR, GOYA, VVDS F2), and with
deeper pencil beam surveys (Hawaii HDFN, WHT, HDFN, HDFS).
}
\label{fig:lognsu}
\end{figure}

Resulting raw counts are shown in Fig.\ref{fig:lognsu} where a clear
decrease is apparent for $U(Vega)>26.4$. The typical photometric
error at $U\sim 27$ is $\sigma=0.4$ magnitude, corresponding to a total
integrated number counts of $\sim$100 galaxies per $arcmin^2$.  An
estimate of the completeness level should be performed in order to
evaluate the amount of correction to the raw counts at the faint
limits. This has been evaluated including in the real image 1000
simulated galaxies per 0.25 magnitude bin in the magnitude
interval $U(Vega)=24-28$ using the standard ''artdata'' package in
IRAF. We include disk galaxies of half light radius of 0.2-0.4 arcsec,
convolved with the PSF of stellar objects in the field. We then
run SExtractor on this new image using the same detection parameters
described above, and we studied how effective SExtractor is in
recovering the simulated galaxies as a function of magnitude. The
resulting sizes of the simulated galaxies are typical of real galaxies
in the magnitude interval $U(Vega)=24-25$, so the completeness we have
computed at $U=26-27$ can be considered a robust estimate. We have
verified using different datasets, i.e. \cite{windhorst} and
GOODS-South \cite{music},
that, at $B\ge 25$, the half light radius of these galaxies is always
between 0.2 and 0.4 arcsec. We choose not to use real sources to
simulate the completeness of the images mainly for two reasons: the
half light radii of galaxies decrease with the magnitude, and it is
known that fainter galaxies are smaller. If we use bright ($U\le 24$) high
signal-to-noise objects, their size, once dimmed to $U\sim 26-27$,
can be overestimated with respect to
their actual size, enhancing artificially the completeness correction
at fainter magnitudes. If we use real faint sources as input in our
simulations, the signal-to-noise ratio of these galaxies is
very low, enhancing even in this case the completeness correction. We
have tried instead a robust and realistic assumption, that is to
simulate galaxies with half light radius between 0.2 and 0.4 arcsec,
which is the range observed at the faint magnitude limits where
morphological analysis is still reliable (see e.g. \cite{conselice}).
The completeness at $U\le 25.5$ does not depend on the half light radius
of the simulated galaxies, while at fainter magnitudes it depends on
the size of the galaxies: in particular, at $U=26.0$ the scatter of
the completeness for various half light radii between 0.2 and 0.4
arcsec is 3\%, increasing to 6\% at $U=26.5$ and reaching 16\% at
$U=27.0$.
The resulting 50\% completeness level is measured
at $U(Vega)=26.6$, while at $U=27.0$ we have a formal completeness of
30\%.

The number counts corrected for incompleteness are shown again in
Fig.\ref{fig:lognsu}. Given the wide magnitude interval from
$U(Vega)=19.5$ to $U(Vega)=27.0$ available in the present survey, the
shape of the counts can be derived from a single survey in a
self-consistent way, possibly minimizing offsets due to systematics in
the photometric analysis of data from multiple surveys (zero point
calibration, field to field variation, etc). A clear bending is
apparent at $U(Vega)> 23.5$. To quantify the effect we fitted the
shape of the counts in the above magnitude interval with a double
power-law. The slope changes from $0.58\pm 0.03$ to $0.24\pm 0.05$ for
magnitudes fainter than $U_{break}=23.6$. The uncertainty in the
break magnitude is however large, $\sim 0.5$, since the transition
between the two regimes of the number counts is gradual.

In Fig.\ref{fig:lognsu} we compare our galaxy number counts with those
derived by shallow surveys of large area (SDSS EDR, \cite{sdss}) or
of similar area (GOYA by \cite{goya};
VVDS-F2 by \cite{radovich}), and with deep pencil beam surveys
(Hawaii HDFN by \cite{capak04}; WHT, HDFN, and HDFS by \cite{wht}). In
particular, the WHT galaxy counts (\cite{wht}) are based on a 34h
exposure time image reaching $U(Vega)=26.8$ but at the much lower
3$\sigma$ level in the photometric noise and in an area of $\sim$50
arcmin$^2$, while the GOYA survey at the INT telescope is complete at
50\% level at $U(Vega)=24.8$. These counts are shown together with the
two pencil beam surveys in the Hubble Deep Fields (\cite{wht}).
The agreement with the GOYA survey (900 sq. arcmin.) is remarkable,
and suggests that once big areas of the sky are investigated, the
effects of cosmic variance are slightly reduced.
The present UV counts obtained during the commissioning of
LBC-Blue are thus a unique combination of deep imaging in the U band
and of large sky area, with the result of a considerable reduction of
the cosmic variance effects for $U\ge 21$. For brighter magnitude limits,
we refer to larger area surveys, shallower than our survey, as shown in
Fig.\ref{fig:lognsu}.

Table \ref{tab:lognsU} summarizes the galaxy number counts, corrected for
incompleteness, with their upper and lower 1 $\sigma$ confidence level
uncertainties, assuming Poisson noise and cosmic variance effect.
For the latter, we used the Cosmic Variance Calculator (v1.02)
developed by \cite{trenti} using as input values a linear size of 22 arcmin
(corresponding to our deeper area of 478.2 $arcmin^2$) and redshift from
0.0 to 3.0, with standard $\Lambda$-CDM cosmology. At U=27, for example,
the computed cosmic variance is 4.5\%.

Using the Q0933+28 and the other three fields in the SXDS area, we study
the field to field variation of the number counts in the U band.
We find that the typical variation from one LBC field to another is 0.04
in LogN for $U\sim 20$, while it reduces gradually to 0.01 at U=24.5,
well below the poissonian uncertainties described in Table \ref{tab:lognsU}.
This ensures that the zero-point calibration of the images is robust
and the area of this survey is sufficient to decrease the cosmic variance
effects below the statistical uncertainties of the galaxy number counts.
We complement these internal checks on the galaxy number counts in U with
an external consistency test, comparing our final number counts with shallower
NC derived in different areas or deep pencil beam surveys in
Fig.\ref{fig:lognsu}. We find that the survey-to-survey maximal variations
are of the order of 0.1 in LogN at all magnitudes, with lower scatter
for the wider surveys considered here (GOYA, SDSS-EDR, VVDS-F2).

\begin{table}
\caption{LBC U galaxy number counts.}
\scriptsize
\begin{tabular}{ccccc}
\hline
\hline
\textbf{U(Vega)} & \textbf{LogN} & \textbf{Max LogN} & \textbf{Min LogN} &
\textbf{Completeness} \\
\hline
19.50 &   2.409 &      2.648 &      1.837 &     1.00 \\
19.75 &   2.484 &      2.707 &      2.001 &     1.00 \\
20.00 &   2.517 &      2.734 &      2.066 &     1.00 \\
20.25 &   2.891 &      3.043 &      2.654 &     1.00 \\
20.50 &   2.991 &      3.129 &      2.787 &     1.00 \\
20.75 &   3.103 &      3.227 &      2.930 &     1.00 \\
21.00 &   3.239 &      3.346 &      3.095 &     1.00 \\
21.25 &   3.350 &      3.446 &      3.227 &     1.00 \\
21.50 &   3.503 &      3.585 &      3.401 &     1.00 \\
21.75 &   3.655 &      3.725 &      3.572 &     1.00 \\
22.00 &   3.782 &      3.843 &      3.711 &     1.00 \\
22.25 &   3.914 &      3.989 &      3.824 &     1.00 \\
22.50 &   4.100 &      4.161 &      4.028 &     1.00 \\
22.75 &   4.191 &      4.247 &      4.127 &     1.00 \\
23.00 &   4.311 &      4.360 &      4.256 &     1.00 \\
23.25 &   4.452 &      4.494 &      4.405 &     1.00 \\
23.50 &   4.561 &      4.598 &      4.521 &     1.00 \\
23.75 &   4.633 &      4.667 &      4.595 &     1.00 \\
24.00 &   4.702 &      4.734 &      4.668 &     1.00 \\
24.25 &   4.799 &      4.827 &      4.768 &     0.99 \\
24.50 &   4.865 &      4.892 &      4.837 &     0.98 \\
24.75 &   4.945 &      4.970 &      4.919 &     0.96 \\
25.00 &   5.016 &      5.040 &      4.992 &     0.93 \\
25.25 &   5.081 &      5.103 &      5.058 &     0.90 \\
25.50 &   5.135 &      5.156 &      5.113 &     0.88 \\
25.75 &   5.199 &      5.219 &      5.178 &     0.83 \\
26.00 &   5.268 &      5.288 &      5.248 &     0.75 \\
26.25 &   5.341 &      5.361 &      5.321 &     0.66 \\
26.50 &   5.431 &      5.450 &      5.411 &     0.55 \\
26.75 &   5.469 &      5.510 &      5.425 &     0.43 \\
27.00 &   5.495 &      5.563 &      5.419 &     0.30 \\
\hline
\hline
\end{tabular}
LBC U galaxy number counts corrected for incompleteness; $N$ is
the number of galaxies per $deg^2$ and per magnitude bin, while the
minimum and maximum counts are the 1 $\sigma$ confidence level (Poisson noise
and cosmic variance effect).
\label{tab:lognsU}
\end{table}

\subsection{The UV extragalactic background light}

The slope of the galaxy number counts in the U band at $U\sim 23.5$
changes from 0.58 to 0.24, which implies that the
contribution of galaxies to the integrated EBL in the UV has a maximum
around this magnitude.  The contribution of observed galaxies to the
optical extragalactic background light (EBL) in the UV band can be
computed directly by integrating the emitted flux multiplied by the
differential number counts down to the completeness limit of the
survey.

Following the work of \cite{mp2000}, we compute the EBL in the U band,
$I_{\nu}$ measured in $erg~s^{-1}~cm^{-2}~Hz^{-1}~Sr^{-1}$, using the
following method:

\begin{equation}
I_{\nu}= 10^{-0.4(U_{AB}+48.6)}N(U_{AB}) \ ,
\end{equation} 

where we use the following correction from Vega to AB:
$U_{AB}=U_{Vega}+0.86$.

\begin{figure}
\includegraphics[width=9cm]{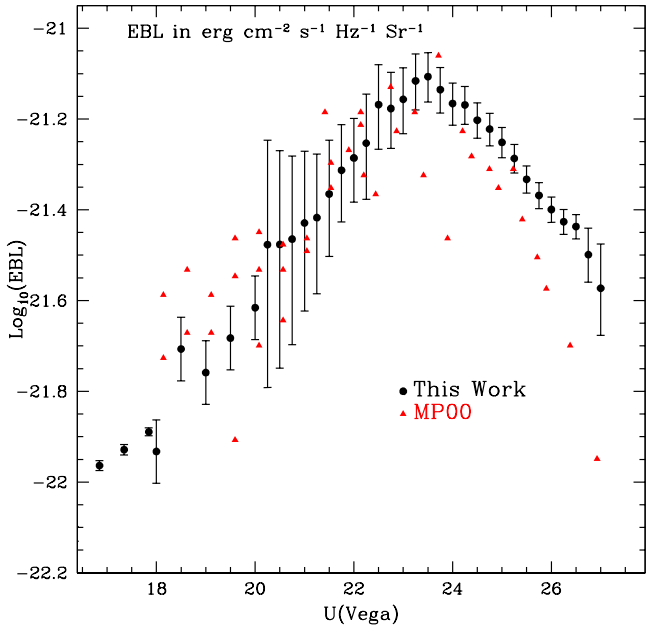}
\caption{
The extragalactic background light per magnitude bin $u_{\nu}$ as a function
of U band magnitudes. We compare the results of this work with the EBL derived
by \cite{mp2000} in the HDFS field. The small error bars of EBL at $U\le 18$
are due to the small uncertainties on the galaxy number counts derived
using SDSS data.
}
\label{fig:ebl}
\end{figure}

Fig.\ref{fig:ebl} shows the EBL obtained by the LBC deep number counts
and compare it with the results of \cite{mp2000}. Our data agrees well
with previous estimates of the EBL in the U band, and indeed it allows
to derive precisely the peak of the differential EBL, at
$U=23.55\pm0.25$, and put strong constraints to the contributions of
faint galaxies down to $U=27.0$ to the integrated EBL in this band.
At magnitudes $U\ge 24$ our estimate of the EBL is significantly
larger than the one of \cite{mp2000}. Checking in detail
Fig.\ref{fig:lognsu}, it is clear that the HDFS is slightly underdense
with respect to the HDFN or other deep pencil beam surveys, probably
due to cosmic variance effects, and this is reflected on the smaller
EBL of \cite{mp2000} compared to the LBC estimate.

The integrated galaxy contribution to the EBL in the UV band results
$\nu I(\nu)=3.428\pm 0.068 nW/m^2/Sr$ at the effective wavelength of
the LBC U-BESSEL filter, 3590\AA, taking into account the magnitude
range U=17-27. This estimate is 20\% higher than the one estimated by
\cite{mp2000} ($2.87^{+0.58}_{-0.42}nW/m^2/Sr$) although their value
is still consistent with our measure due to their large
uncertainties. Moreover, our estimate reduces the uncertainties by an
order of magnitudes, which is of fundamental importance
to put strong constraints to galaxy
evolution models. Extrapolating the observed number counts in the U
band down to flux=0 we derive an integrated EBL of
$\nu I(\nu)=3.727\pm 0.084nW/m^2/Sr$,
so in our survey to U=27 we are resolving $92\%$ of EBL produced by
galaxies.

The integrated flux from resolved galaxies, however, should be considered
as a lower limit to the total EBL in the universe, since our
estimate could be affected by different systematics:

\begin{itemize}
\item
Incompleteness in the number counts at faint magnitudes due to the presence
of very extended/diffuse sources that escape detection in our LBC deep images
and in the HDFs surveys. The adopted detection algorithms usually select
galaxies down to a threshold in their surface brightness limits, and so
are prone to selection effects against low surface brightness galaxies.
We have computed the completeness of galaxies assuming a conservative
half light radius of 0.3 arcsec, and we cannot exclude the presence
of a population of diffuse and extended galaxies in the UV bands, even
this hypothesis is quite unlikely, due to the well known relation between
observed optical magnitudes and half light radius of galaxies
(\cite{totani01}).
\item
The surface brightness dimming of sources at high redshifts. This is
not a serious problem in the U band, since all the light in this
wavelengths comes from $z\le 3$ galaxies (non U-dropout galaxies) and,
as it is discussed in the next paragraph, the redshift distribution of galaxies
contributing to the peak of the EBL are at $z\le 1$.
\end{itemize}

Independent estimates to the total EBL in the U band are given in
\cite{bernstein02,bernstein07}, and are slightly larger
than our estimate from the galaxy number counts
($\nu I(\nu)=14.4\pm 9 nW/m^2/Sr$ in \cite{bernstein02}
and $\nu I(\nu)=(21.6\pm 14.4) nW/m^2/Sr$ in \cite{bernstein07},
after a different treatment for the contaminating foregrounds).

With the present observations, we reject the hypothesis of
\cite{bernstein02} that a huge contribution to the sky background
in the U band could be derived by the overlapping wings of extended
galaxies. With the present observations we reach a surface brightness limit
of $U=28.2 mag/arcsec^2$ at 3$\sigma$ and thus exclude the presence of
faint and extended wings in the galaxy UV light distributions.

The discrepancy between the resolved and total EBL could be explained
by a population of ultra faint and numerous galaxies at $U\ge 27$,
beyond the current limits of the present surveys and with a LogN-LogS
slope steeper than 0.5, or by an improper subtraction of the
foreground components, as stated in \cite{mattila03} and \cite{bernstein07}.

Recently, \cite{fornax} proposed a new method to derive stringent
limits to the diffuse light in the UV bands, using the inverse compton
emission in $\gamma$-rays of the EBL by the high energy electrons in
the radio lobes of Fornax A.  This technique will allow to derive
constraints to the EBL at shorter wavelengths than the limits obtained
by the TeV blazar emission (\cite{stanev,aharonian,albert}). At the
present stage, \cite{albert} give an upper limit of $\nu I(\nu)\le 5
nW/m^2/Sr$ to the total EBL in the UV. This limit strengthens the
hypothesis that the discrepancy between our estimate of the EBL in U
band and that of \cite{bernstein07} is due to a foreground local
component.

Our derivation of the EBL by integrated galaxy counts is marginally consistent
with the lower limit derived by \cite{kd08} of $\nu I(\nu)\ge 3.97
nW/m^2/Sr$ at $\lambda=3600$\AA.


\section{Comparison with theoretical models}

The availability of galaxy number counts down to faint magnitude
limits and over large sky areas can be used to test the predictions of
different theoretical models without being strongly
affected by the cosmic
variance. Among the various models (numerical and
semi-analytical) developed in the framework of the
standard hierarchical CDM scenario, we have selected the models
developed by \cite{menci06} (hereafter M06),
the \cite{kw07} model based on the Millennium Simulation (hereafter
K07) and the model developed by \cite{morgana} (hereafter {\sc MORGANA}).
All these model simulate
the formation and evolution of galaxies, starting from a statistical
description of the evolution of the DM halo population (merger trees),
and using a set of approximated, though physically motivated, ``recipes''
to treat the physical processes (gas cooling, star formation, AGN and SF
feedback, stellar population synthesis) acting on the barionic component.
An appropriate treatment of
dust attenuation is also crucial in comparing model predictions to the
UV observations, given the efficient scattering of radiation at
these wavelengths by dust grains, as shown by GALEX number counts of
\cite{xu2005}.

The comparison of the observed counts with that predicted by the three
selected models is used to enlighten different critical issues
concerning the physical description of the galaxy formation and
evolution.

\begin{figure}
\includegraphics[width=9cm]{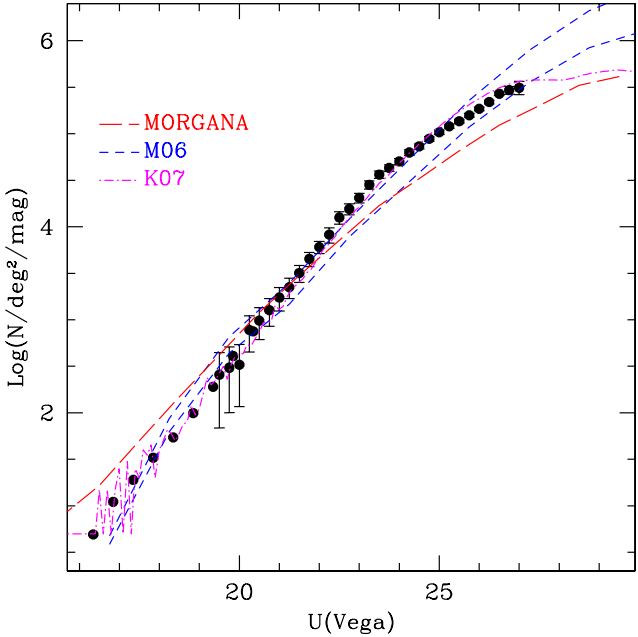}
\caption{
Number counts of galaxies in the U-BESSEL band for the Q0933+28 LBC
field, complemented at brighter magnitudes ($U\le 20$) by the number
counts of VVDS-F2 and SDSS-EDR, whose error bars are always smaller
than the size of the points.
We compare our counts with theoretical models (M06, K07, {\sc MORGANA}).
}
\label{fig:models}
\end{figure}

In particular, the M06 model is characterized by a specific
implementation of the AGN feedback on the star formation activity in
high redshift galaxies. The description is based on the expanding
blast waves as a mechanism to propagate outwards the AGN energy
injected into the interstellar medium at the center of galaxies (see
Menci et al. 2008); such a feedback is only active during the active
AGN phase (``QSO mode''), and it is effective in suppressing the star
formation in massive galaxies already by $z\approx 2$, thus yielding a
fraction of massive and extremely-red objects in approximate agreement
with observations (Menci et al. 2006).
The M06 model uses three different prescriptions
for the dust absorption, namely the Small Magellanic Cloud model (SMC),
the Milky Way (MW) law and the \cite{calzetti} extinction curve (C00).
In Fig.\ref{fig:models} the two short-dashed curves enclose the minimum
and maximum number counts predicted by the M06 model taking into account the
three different extinction curves (SMC, MW, and C00).

The model of \cite{kw07} is based on the Millennium Simulation
(\cite{springel}), on a concordance Lambda CDM cosmology. In the
model an important critical issue is represented by the particle mass
resolution of the code, which affects the physical and statistical
properties of faint low-luminosity galaxies. The adopted CDM mass
resolution is sufficient to resolve the halos hosting galaxies as faint
as 0.1 $L^*$ with at least 100 particles of $8.6 \cdot 10^{8} h^{-1} M_\odot$.
The model also adopts a different AGN feedback called
"radio mode" in which cooling is suppressed by the continuous
accretion of hot gas onto Supermassive Black Holes at the center of
groups or cluster of galaxies.
The only difference of K07 from \cite{croton06} and \cite{dlb07}
is in the
dust model used in their simulations. For the local galaxies, they adopt
a simple relationship between face-on optical depth and intrinsic luminosity
$\tau\propto \tau_0 (L/L_*)^\beta$ with $\beta=0.5$. At high redshifts
they take into account the different dust and gas contents, the varying
metallicities, and the shorter rest frame emitted wavelengths. In particular
their average extinction increases strongly at high redshifts due to
the smaller disc sizes of the galaxies. We refer to \cite{kw07} for
details about their dust model.

Finally, the {\sc MORGANA} model is characterized by a different
treatment of the processes of gas cooling and infall (following Viola
et al., 2008), star formation and feedback (using the multi-phase
model of Monaco 2004). Black hole accretion is described in detail in
Fontanot et al., (2006): the accretion rate in Eddington units
determines the nature of feedback from the AGN. Synthetic SEDs for
model galaxies are obtained using the GRASIL code (Silva et al. 1998),
which explicitly solves the equation of radiative transfer in a dusty
medium. Fontanot et al. (2007) show that {\sc MORGANA} is able to
reproduce both sub-mm ($850 \mu$m) number counts and the redshift
distribution observed in $K$-limited samples, with conservative choice
for the stellar IMF. Galaxy SEDs, magnitudes, and colours for a
variety of passbands are obtained using the {\sc GRASIL}
spectrophotometric code (Silva et al.  1998), which explicitely solves
the equations for the radiative transfer in a dusty medium taking into
account the composite geometry (bulge+disc) of each model galaxy. The
dust properties (dust-to-gas mass fractions, composition, and size
distribution) are kept fixed to values providing a good agreement with
local observations (see \cite{ff07}, for more details on the coupling
between {\sc MORGANA} and {\sc GRASIL}). In this work we take
advantage of the recent update of the model, adapted for WMAP3
cosmology and Chabrier IMF and presented in Lo Faro et al. (2009).

It is
worth noticing that photoelectric absorption by the interstellar
medium is redshifted, for high redshift galaxies, into the observed UV
band. For this reason only galaxies with $z<3$ contribute to the UV
number counts. The common characteristics of the models is that
only galaxies in the $1.5<z<2.5$ redshift intervals give the
main contribution to the counts at $U>26$ (see \cite{barro2009}).
This implies that the shape of the
faint counts is provided by the average faint-end shape of the galaxy
luminosity function in the same redshift interval.

The comparison of the observed galaxy number counts in the U
(360 nm) band with the predictions of
theoretical models shown in Fig.\ref{fig:models} indicates that:
\begin{itemize}
\item
the M06 and K07 models reasonably reproduce the UV number counts from
$U=17$ to $U=27$, while the {\sc MORGANA} model shows the largest deviation
from the observed data.
\item
The three models show different shapes for the UV galaxy number counts
at faint fluxes, implying discrepancies which increase for increasing
magnitudes. This can be evaluated in more detail dividing the counts
in redshift bins (see \cite{barro2009}).
\item
With the current observations of the UV number counts down to $U=27$
both M06 and K07 models show a similar behaviour, although the K07
model appears to bend over for $U>27$ in contrast with the steeper
slope of the M06 model.
\end{itemize}

The discrepancies between predicted and observed faint UV galaxy
number counts could be due in principle to several effects, e.g. the
evolution in number density, star formation activity or
dust extinction of the faint galaxy population. Among these, an
appropriate treatment of the dust extinction plays an important role,
in this comparison based on the U band, since it can affect both the
normalization and shape of the observed UV counts. The effects due to
the different extinction laws (Calzetti 2000 and Small Magellanic
Cloud) adopted in the M06 model for example, increase at fainter
magnitudes since we are observing galaxies at high ($z<3$)
redshifts. In this respect, the bending observed in the UV counts
could be indicative of a gradual redshift evolution of the dust
properties.

Another physical quantity relevant for the correct reproduction of the
faint counts is the amount of star formation activity in faint low-mass
galaxies. This effect is particularly important in the MORGANA model
where it is responsible for the flat shape of the predicted counts.
\cite{fontanot09} show that the underprediction of UV number counts of
faint galaxies can be explained by a rapid decline in their star
formation activity and consequently in their associated UV emission at
$z\lesssim 2$. These galaxies are predicted to be too passive and to host
too old stellar populations at later times with respect to observations.
The reduced SFRs can easily explain the flat shape of the predicted
counts in {\sc MORGANA}.

\begin{figure}
\includegraphics[width=9cm]{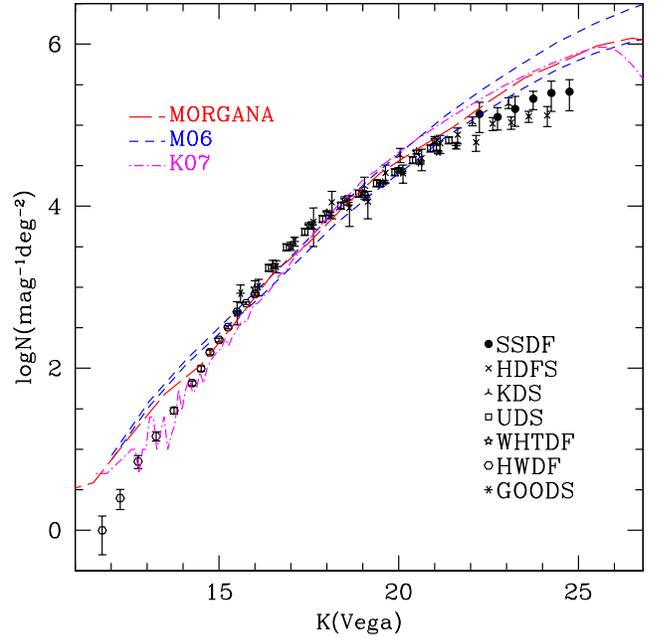}
\caption{
Number counts of galaxies in the K band for different deep surveys
in the literature.
The model predictions come from the same models described in the previous 
figure.
}
\label{fig:lognsk}
\end{figure}

As an attempt to disentangle dust/SFR effects from
number density evolution, we have extended the model comparison to the
NIR K band number counts which are much less sensitive to dust
absorption and short episodes of star formation.
Indeed, the observed flux in the K band is more related
to the star formation history in the galaxy quantified by its
assembled stellar mass (see e.g. Fontana et al. 2006).
Fig.\ref{fig:lognsk} shows the comparison of the observed number
counts in the K band collected in the literature and provided by
different instruments/telescopes and surveys (SSDF: \cite{ssdf}, HDFS:
\cite{hdfs}, KDS: \cite{kds}, UDS: \cite{uds}, WHTDF: \cite{whtdf},
HWDF: \cite{hwdf}, GOODS: \cite{music}) with the predictions of M06,
K07, and {\sc MORGANA}. The observed K band galaxy counts show a clear
bending at $K\simeq 17$ where the average slope changes from 0.69 to a
flatter value 0.33. Here the bending is more clear compared to what
observed in the UV since the break magnitude is brighter with respect
to the faintest survey limits.

In the K band the models tend to overestimate the number counts both
at the bright end $K<15$ and at the faint magnitudes $K>20$. The
excess in the predicted K band number counts resembles that at the
faint end of the theoretical luminosity functions in the NIR (see for
example \cite{poli} and \cite{fontanot09}). This can be indicative of
an excess in the production of galaxies with low luminosity and high
redshifts. If the NIR number counts are extrapolated at fainter
limits, the amount of this excess differs among the models, being
smaller in the K07 model at $K\ge 26$. The latter however is affected
by a threshold in the mass resolution used
by the numerical code of the Millennium simulation adopted in the K07
model. This could provide an artificial removal of faint galaxies.

The simultaneous comparison of the UV and NIR K band number
counts at faint fluxes seems to indicate that the underprediction of
the modelled UV number counts could be due to dust extinction or
reduced SF activity rather than to an intrinsic evolution in the
galaxy number density.

Finally we identify a third issue when comparing the bright end of the
galaxy counts. The treatment of the feedback on SF processes due to
AGN activity can play an important role in this respect. The M06 model
for example shows the effect of a different treatment of AGN feedback
based on the so called QSO mode at variance with the Radio mode
feedback used by the K07 model, while {\sc MORGANA} is characterized
both by Radio and by QSO modes. The different treatment of AGN feedback is
plausibly the responsible for the differences in the predictions of
NIR galaxy number counts at $K\le 15$. The excess of the UV number
counts predicted by models is related to the quenching of the star formation
activity. {\sc MORGANA} tend to
overpredict star formation in massive central galaxies at low-z. This is
related to the less efficient, or delayed, quenching of the cooling
flows in massive halos via Radio-mode feedback. In fact, in this model
gas accretion onto the central black hole is related to star formation
activity in the spheroidal component: this implies that AGN heating
switches on only after some cooled gas has already started forming stars
in the host galaxy (see Kimm et al., 2008 for a complete discussion and
a comparison of different Radio-mode implementations in semi-analytical
models).

As a last comment, all the three models give a contribution to the UV
EBL which is broadly consistent with the upper limits described in section
3.2. Using the same method described there for the observed
data, we derived an EBL of 0.71-1.18 $nW/m^2/Sr$ for the M06 model in
the UV band, while the {\sc MORGANA} and K07 ones give 2.66 and 3.26
$nW/m^2/Sr$, respectively.

\section{Summary}

To summarize the main results of the paper:
\begin{itemize}
\item
We have derived in a relatively wide field of 0.4 deg$^2$ the deepest
counts in the 360nm UV band. This allowed to evaluate the shape of the
galaxy number counts in a wide magnitude interval U=19-27
with the advantage of mitigating the cosmic variance effects.
The agreement with the number counts of shallower surveys confirms
the low impact of systematic errors on the LBC galaxy statistics.
\item
The shape of the counts in UV can be described by a double power-law
with a steep slope 0.58 followed beyond $U\simeq 23.5$ by a flatter
shape 0.24. Our counts are consistent at the bright end with surveys
of comparable or greater areas. At the faint end our counts are more
consistent with that found in the HDF-N.
\item
The faint-end slope of the counts is below 0.4 and this ensures
the convergence of the
contribution by star forming galaxies to the EBL in the UV band. The
total value in the UV band obtained extrapolating the slope of our
counts to flux=0 is indeed $3.727\pm 0.084 nW/m^2/Sr$.
It is consistent with recent upper limits coming from TeV
observations of \cite{albert}, $\nu I(\nu)\le 5 nW/m^2/Sr$,
showing that the UV EBL is resolved at $\ge$74\% level.
\item
We have compared our counts in the UV and K bands with few selected
hierarchical CDM models which are representative of specific critical
issues in the physical description of the galaxy formation and
evolution.
\item
The mass resolution of numerical models is critical for reproducing the
faint end of the UV galaxy number counts.
\item
The discrepancies between predicted and observed UV galaxy number counts
at faint magnitudes could be mainly due to the treatment of dust extinction and
the star formation activity in low mass galaxies at $z\le 2$.
\item
The AGN feedback (Radio vs QSO mode) may affect galaxy counts at the bright
end of the LogN-LogS in the K band.
\end{itemize}

A correct physical description of the AGN feedback, dust properties
and star formation activities in the models is fundamental to ensure a
reasonable agreement of the model predictions at the faint end of the
galaxy counts.

Adding colour information for galaxies with UV emission as faint as
$U=27-28$ implies very deep observations in the red bands which are
feasible with several hours of integration at 8m class telescopes.
Very deep multicolour information on areas of the order of the square
degree can help in extracting physical information on the star
formation history of the dwarf population at intermediate and high
redshifts.


\begin{acknowledgements}
Observations have been carried out using the Large Binocular Telescope
at Mt. Graham, Arizona, under the Commissioning phase of the Large Binocular
Blue Camera. The LBT is an international collaboration among
institutions in the United States, Italy and Germany. LBT Corporation
partners are: The University of Arizona on behalf of the Arizona
university system; Istituto Nazionale di Astrofisica, Italy; LBT
Beteiligungsgesellschaft, Germany, representing the Max-Planck
Society, the Astrophysical Institute Potsdam, and Heidelberg
University; The Ohio State University, and The Research Corporation,
on behalf of The University of Notre Dame, University of Minnesota and
University of Virginia. The Millennium Simulation databases used in
this paper and the web application providing online access to them
were constructed as part of the activities of the German Astrophysical
Virtual Observatory.
Some of the calculations were carried out on the PIA cluster of the
Max-Planck-Institut f\"ur Astronomie at the Rechenzentrum Garching.
We thank the anonymous referee for useful comments which helps in improving
the quality of the present paper.
AG warmly thanks Kalevi Mattila and Martin Raue for useful comments on the
EBL limits.
\end{acknowledgements}

\end{document}